\documentclass[10pt,conference]{IEEEtran}\IEEEoverridecommandlockouts
\usepackage{cite}
\usepackage{amsmath,amssymb,amsfonts}
\usepackage{algorithmic}
\usepackage{graphicx}
\usepackage{textcomp}
\usepackage{xcolor}
\usepackage{hyperref}
\def\BibTeX{{\rm B\kern-.05em{\sc i\kern-.025em b}\kern-.08em
    T\kern-.1667em\lower.7ex\hbox{E}\kern-.125emX}}
\begin{document}

\title{Shadow-Based Noise Fingerprinting of Simulated Quantum Noise Models}

\author{\IEEEauthorblockN{Vridhi Jain}
\IEEEauthorblockA{\textit{Department of Quantum Science and Engineering} \\
\textit{University of Delaware}\\
DE, USA \\
vridhi@udel.edu}
\and
\IEEEauthorblockN{Lei Zhang}
\IEEEauthorblockA{\textit{Department of Information Systems} \\
\textit{University of Maryland, Baltimore County}\\
MD, USA \\
leizhang@umbc.edu}
}

\maketitle

\begin{abstract}

Accurate noise classification is essential for operating near-term quantum processors, yet existing approaches, such as quantum process tomography, scale exponentially with system size, limiting their practicality for routine calibration. We propose a measurement-efficient noise fingerprinting pipeline that combines structured classical shadow tomography with physics-informed feature engineering to identify noise channels from a fixed set of 3-qubit probe circuits. Each sample is represented by a feature vector constructed from randomized Pauli measurements and derived observables designed to resolve physically similar noise channels that produce overlapping signatures under generic measurement sets. We evaluate random forest, extra trees, and a multilayer perceptron on 10,000 labeled samples spanning ten noise models. The three classifiers achieve comparable performance. In the reported runs, random forest and extra trees perform similarly, achieving approximately 0.736 test accuracy and 0.729--0.730 macro F1, compared with 0.715 accuracy and 0.699 macro F1 for the multilayer perceptron. We further analyze the effect of the noise-strength sampling range and conduct a limited sensitivity check using analogous 2- and 4-qubit probes. Confusion analysis shows that readout error, phase flip, thermal relaxation, and bit flip are classified with high reliability, while most remaining errors occur among channels with similar physical effects.

\end{abstract}


\section{Introduction}
Quantum processors based on superconducting qubits and trapped ions are rapidly advancing toward practical applications, yet their performance remains fundamentally constrained by hardware noise~\cite{preskill2018quantum}. Noise in these systems arises from a variety of physical mechanisms, including energy relaxation, dephasing, and measurement errors, each of which degrades quantum information in distinct ways. Understanding the dominant noise sources in a given device is therefore a prerequisite for effective error mitigation and the design of noise-aware quantum algorithms.

Existing approaches, such as quantum process tomography~\cite{nielsen2010quantum}, fully characterize noise processes but require resources that scale exponentially with system size, making them impractical for routine calibration of large-scale devices. Direct fidelity estimation and  randomized benchmarking offer partial improvements in scalability, yet  remain limited to specific noise models or rely on hardcoded decision  rules that do not generalize across diverse error channels. As processors scale in qubit count and circuit depth, efficient and generalizable noise identification becomes increasingly critical~\cite{preskill2018quantum}.

In this work, we address this gap with a noise fingerprinting pipeline\footnote{Here, we use the term ``noise fingerprinting'' to refer to supervised classification of predefined noise-model families from shadow-derived measurement features. Our study focuses on simulated Qiskit noise channels and does not attempt full noise characterization or continuous noise-parameter estimation.} that uses classical shadows~\cite{huang2020predicting} and physics-informed features extracted from nine 3-qubit probe circuits~\cite{farhi2014quantum}. Across ten simulated noise models, random forest, extra trees, and the multilayer perceptron (MLP) achieve comparable benchmark performance, with the tree-based classifiers performing slightly better in the reported runs. However, these relatively small differences may vary with the random seed and generated dataset.

Our \textbf{contributions} are threefold. First, we propose a shadow-based noise fingerprinting pipeline for 3-qubit probe circuits, using a physics-informed feature representation that combines Pauli-shadow observables with derived coherence, population, and asymmetry features. Second, we evaluate three machine-learning classifiers across ten simulated Qiskit noise models and analyze the channel pairs that remain difficult to distinguish. Third, we characterize two sources of experimental sensitivity: 1) the range from which noise strengths are sampled in the primary 3-qubit setting and 2) the use of analogous 2- and 4-qubit probe circuits. Our artifacts are publicly available at \url{https://github.com/AdaHgrace/Noise_Fingerprinting}.

\section{Background and Related Work}\label{sec:background}

\textbf{Noise models.} Noise in noisy intermediate-scale quantum (NISQ) devices~\cite{preskill2018quantum} arises from distinct physical mechanisms that can be modeled as quantum channels acting on the system's density matrix~\cite{nielsen2010quantum}. We consider ten noise models spanning a broad range of error types: depolarizing, phase flip, bit flip, readout error, phase damping, thermal relaxation, amplitude damping, phase-amplitude damping, Pauli-asymmetric, and reset noise~\cite{nielsen2010quantum}. These range from symmetric decoherence channels to combined energy-and-coherence loss models, providing a diverse and physically realistic benchmark for noise fingerprinting.

\textbf{Shadow tomography.} Shadow tomography predicts properties of a quantum state without requiring complete state reconstruction~\cite{aaronson2018shadow}. Classical-shadow protocols make this approach practical through randomized measurements and compact classical representations~\cite{huang2020predicting,struchalin2021experimental,huang2022learning}. Extensions to noisy settings further support their use under realistic measurement conditions~\cite{koh2022classical}. Their reduced measurement requirements motivate our use of shadow-derived observables as noise-sensitive features.

\textbf{Quantum noise classification and characterization.} Quantum noise has been studied through process tomography, fidelity estimation, randomized benchmarking, statistical fingerprints, and machine-learning analysis of measurement data~\cite{nielsen2010quantum,magesan2011scalable,martina2022learning,canonici2024machine,martina2023deep,mukherjee2024noise}. Recent studies have also investigated efficient randomized-measurement and shadow-inspired representations for characterizing quantum systems and noise~\cite{struchalin2021experimental,hu2022hamiltonian,bensoussan2025toward}. However, prior approaches generally estimate aggregate error properties, distinguish a limited number of noise processes, or learn device-specific signatures. Our work instead uses a common probe and shadow-measurement framework with physics-informed features to classify ten predefined noise-model families.

\section{Our Method}

\begin{figure*}[thbp!]
\centerline{\includegraphics[width=\textwidth]{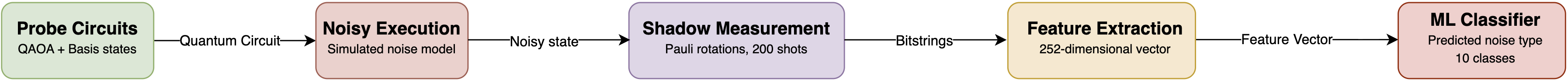}}
\caption{Shadow-based noise-fingerprinting pipeline: probe-state preparation, noisy execution, randomized Pauli measurement, feature extraction, and machine-learning classification.}
\label{fig:pipeline}
\end{figure*}

\subsection{Pipeline overview}

Our noise fingerprinting pipeline maps measurements from fixed 3-qubit probe circuits to one of ten candidate noise models. As illustrated in Fig.~\ref{fig:pipeline}, it consists of five stages: 1) probe-state preparation, 2) simulated noisy execution, 3) randomized shadow measurement using Pauli rotations and 200 shots, 4) extraction of a 252-dimensional physics-informed feature vector, and 5) machine-learning classification. These stages are designed to extract discriminative information with limited measurements. The pipeline is implemented in simulation using Qiskit~\cite{javadi2024quantum} and evaluated on 10,000 labeled samples distributed equally across the ten noise models.

\subsection{Probe state design}
We use 3-qubit QAOA circuits~\cite{farhi2014quantum} as complex probe states, owing to their ability to generate entangled states with rich coherence structure that is sensitive to a broad range of noise mechanisms. QAOA circuits are parameterized by alternating cost and mixer layers, and in this work we use a single-layer ansatz with randomly sampled parameters to ensure diversity across probe instances. These are supplemented with four simple structured probes: 1) the basis state $|0\rangle^{\otimes n}$, 2) the basis state $|1\rangle^{\otimes n}$, 3) the superposition $|+\rangle^{\otimes n}$, and 4) a Greenberger--Horne--Zeilinger (GHZ) state $(|0\rangle^{\otimes n}+|1\rangle^{\otimes n})/\sqrt{2}$ with $n=3$ qubits, prepared via a Hadamard on the first qubit followed by a CNOT chain propagating entanglement to every remaining qubit. The combination of  complex and simple probe states ensures that the feature extraction stage receives complementary information with population-level signatures from the basis states, coherence-level signatures from the superposition state, and correlation-based signatures from the GHZ state, improving the discriminability of physically similar noise channels such as phase damping and thermal relaxation.

\subsection{Feature construction}

For each 3-qubit probe, the observable set contains the single-qubit Pauli observables \(X_i\), \(Y_i\), and \(Z_i\) at every qubit, together with the same-axis 2-qubit correlations \(X_iX_j\), \(Y_iY_j\), and \(Z_iZ_j\) for every qubit pair. This yields 18 raw observables per probe. Each probe circuit is followed by a randomly sampled Pauli-basis rotation, and expectation values are reconstructed from the resulting bit strings over 200 shots using the classical-shadow protocol~\cite{huang2020predicting}.

Each probe also contributes ten derived features built from the mean absolute responses over the three Pauli-axis observable groups. These features summarize axis-specific response, coherence strength, population strength, pairwise asymmetry, and ratios between coherence- and population-sensitive responses. The design targets channels, such as depolarizing and Pauli-asymmetric noise, that can produce nearly indistinguishable signatures under generic observable sets. Across the nine probe circuits, the resulting 3-qubit feature vector has 252 dimensions.

\subsection{Classifiers}

We benchmark three machine learning classifiers on the extracted feature vectors to assess whether model complexity is a limiting factor in noise model identification. The first two classifiers 
are random forest and extra trees. Both methods construct ensembles of decision trees while considering random subsets of features. Random forest ordinarily fits trees to bootstrap samples, whereas extra trees introduces stronger randomization by selecting candidate split thresholds randomly. The third classifier is an MLP with three hidden layers of 256, 128, and 64 units respectively, using ReLU activations and the Adam optimizer, included as a neural baseline to evaluate whether the added expressivity of a neural network provides any advantage over ensemble methods on this task.

All three classifiers are implemented using scikit-learn~\cite{pedregosa2011scikit}, with random forest and extra trees using 300 and 500 trees respectively and class-balanced sample weighting; remaining hyperparameters are left at their scikit-learn default values.

\section{Results}\label{sec:results}

\subsection{Dataset} 

Each dataset consists of 10,000 labeled samples, 1,000 per noise model, generated by applying the fixed nine-probe set to a single noise configuration per sample. The primary error parameter of each model is sampled uniformly from a specified strength range. 
We evaluate three noise-strength ranges in the 3-qubit setting, namely \([0.05, 0.25]\), \([0.1, 0.5]\), and \([0.1, 1.0]\), to characterize the effect of noise strength on classification difficulty. We adopt \([0.1, 0.5]\) as the primary experimental condition. Each dataset is split via stratified sampling into a 20\% test set of 2,000 samples and an 85:15 train/validation split of the remainder, with 6,800 and 1,200 samples, respectively, and equal class representation preserved throughout. For each configuration, classifiers are evaluated across three seeds (42, 43, and 44) controlling the data split and model initialization. We report the mean and standard deviation of test accuracy and macro F1 across these seeds.

\subsection{Experimental results}

\begin{table}[thbp!]
\caption{Classification performance on the test set.}
\begin{center}
\begin{tabular}{|l|r|r|}
\hline
\textbf{Classifier} & \textbf{Accuracy} & \textbf{Macro F1} \\
\hline
Extra Trees & $0.7358 \pm 0.0064$ & $0.7288 \pm 0.0064$ \\
Random Forest & $0.7355 \pm 0.0104$ & $0.7299 \pm 0.0113$ \\
MLP & $0.7150 \pm 0.0115$ & $0.6987 \pm 0.0237$ \\
\hline
\multicolumn{3}{l}{\footnotesize Values reported as mean $\pm$ standard deviation over three random seeds.}\\
\end{tabular}
\label{tab:results}
\end{center}
\end{table}

Table~\ref{tab:results} reports held-out test performance. Extra trees achieved the highest mean accuracy of 0.7358, closely followed by random forest (0.7355), with random forest slightly ahead on macro F1 (0.7299 versus 0.7288). Both tree-based methods perform slightly better than the MLP in these runs, although the relatively small differences should not be interpreted as establishing the general superiority of either classifier family.

\subsection{Confusion analysis}

\begin{figure}[thbp!]
\centerline{\includegraphics[width=\columnwidth]{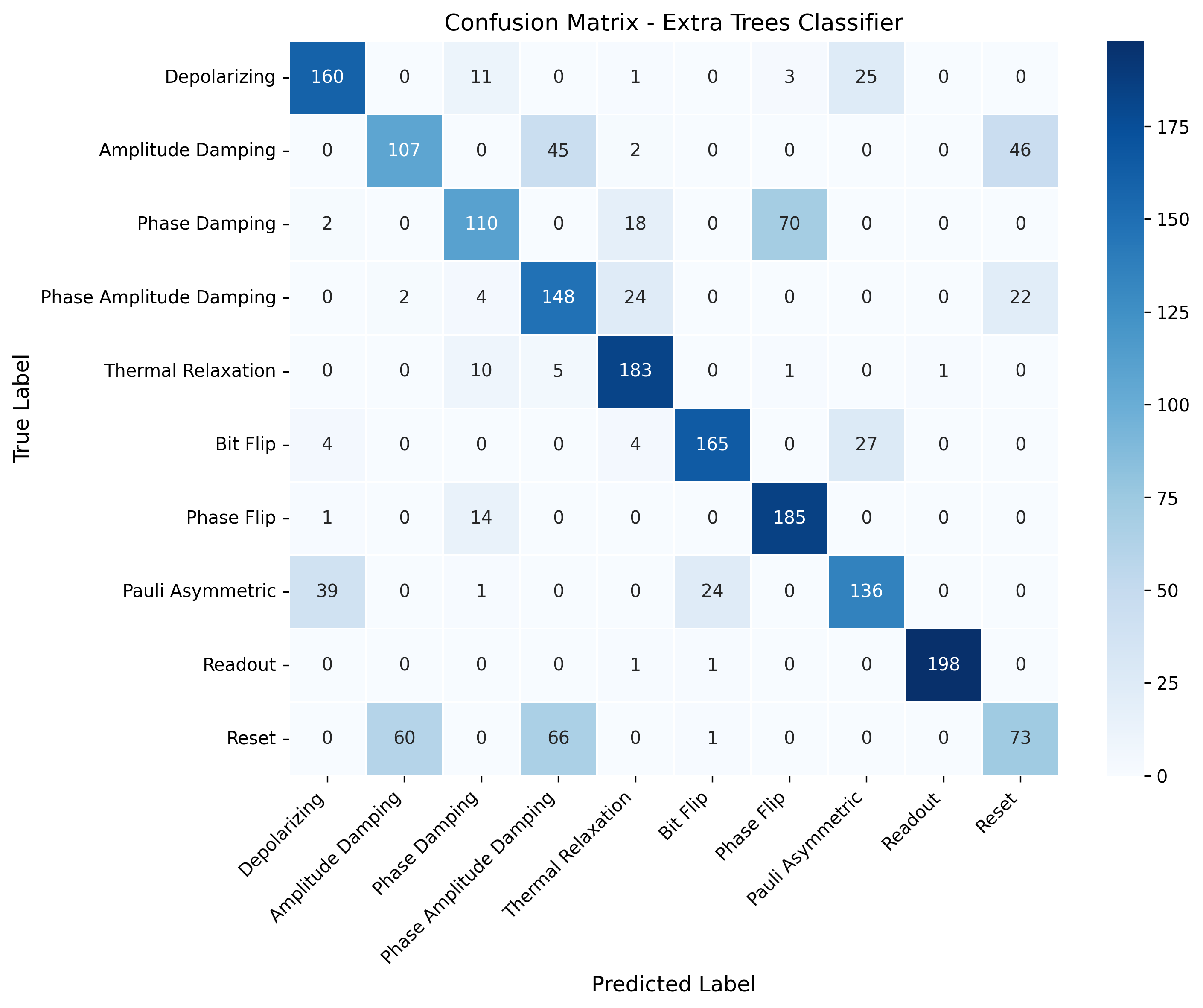}}
\caption{Confusion matrix for the extra trees classifier evaluated on the held-out test set across ten noise models (strength range \([0.1,0.5]\)).}
\label{fig:confusion}
\end{figure}

Figure~\ref{fig:confusion} presents the confusion matrix for the extra trees classifier evaluated on the held-out test set. Readout error, phase flip, bit flip, and thermal relaxation achieve high per-class accuracy, with the classifier correctly identifying the majority of samples in each of these categories. The primary misclassifications occur between phase damping and phase flip, and among amplitude damping, phase-amplitude damping, and reset noise, suggesting that these channel groups produce similar feature signatures under the current measurement protocol. The remaining noise models, including depolarizing and Pauli-asymmetric noise, achieve moderate accuracy, with errors spread across several physically related channels rather than concentrated in one class.

\subsection{Effect of noise strength on classification accuracy}\label{sec:strength-analysis}

\begin{figure}[thbp!]
\centerline{\includegraphics[width=\columnwidth]{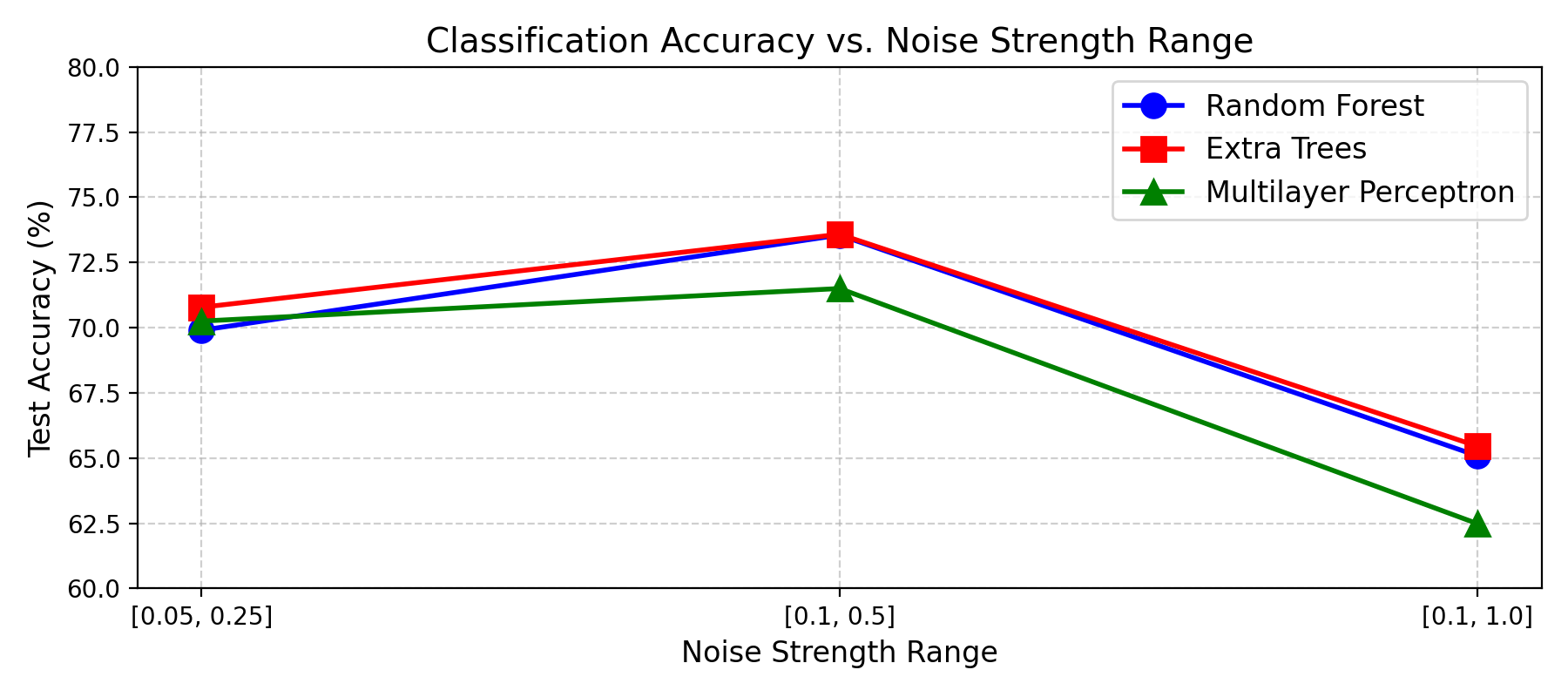}}
\caption{Test accuracy across noise-strength sampling ranges for random forest, extra trees, and MLP classifiers.}
\label{fig:strength}
\end{figure}

Figure~\ref{fig:strength} reports test accuracy at 3 qubits across three noise-strength sampling ranges. Accuracy is highest for the intermediate range $[0.1, 0.5]$ and lower for both the narrower low-strength range $[0.05, 0.25]$ and the widest range $[0.1, 1.0]$, consistently across all three classifiers. This non-monotonic trend suggests two competing effects. At low strength, the noise signal is weak relative to shot noise, making channels harder to distinguish. At the other extreme, sampling strength from a wide range such as $[0.1, 1.0]$ increases within-class feature spread, since the same noise type produces substantially different signatures at low and high strength, which in turn increases overlap between classes. The intermediate range $[0.1, 0.5]$ balances these effects, providing sufficient signal strength while limiting within-class variability, and is adopted as our primary experimental condition.

\section{Discussion}

\subsection{Analysis of classification performance}

The observed classification performance indicates that structured shadow measurements with physics-informed features provide a discriminative representation for several of the ten noise models in the 3-qubit setting. The confusion between phase damping and phase flip is consistent with these channels sharing a dephasing-type mechanism and therefore producing similar expectation-value profiles under the current probes and observables. Likewise, the confusion among amplitude damping, phase-amplitude damping, and reset noise reflects their overlapping effects on population and coherence observables. The three classifiers achieve broadly comparable performance. Although the tree-based classifiers perform slightly better in the reported runs, these results are based on a single generated dataset. Regenerating the simulated dataset may alter the observed differences. Consequently, the results should be interpreted as outcomes of this benchmark rather than evidence that one classifier family is generally superior.

\subsection{Sensitivity to probe-system size}\label{sec:qubit-analysis}

\begin{figure}[thbp!]
\centerline{\includegraphics[width=\columnwidth]{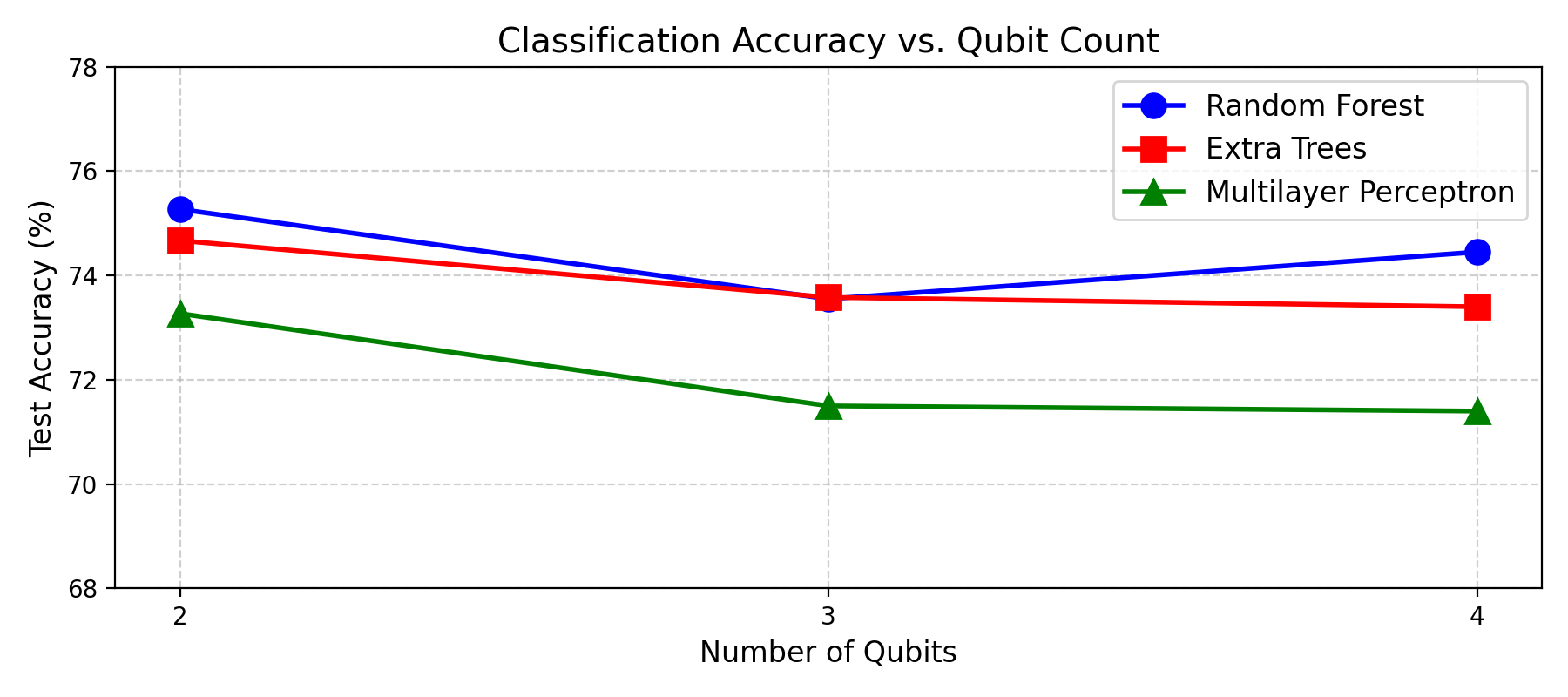}}
\caption{Classification accuracy across probe sizes at the \([0.1, 0.5]\) noise-strength range (mean over three seeds).}
\label{fig:qubit}
\end{figure}

As a secondary robustness check, we repeat the experiment with analogous 2- and 4-qubit probe circuits while holding the noise-strength range at \([0.1, 0.5]\). As shown in Fig.~\ref{fig:qubit}, accuracy ranges from 0.714 to 0.753 across the three classifiers and the three probe sizes. Random forest performs best at two and four qubits while extra trees leads narrowly at three qubits. These results show that performance is relatively stable across the nearby probe sizes considered and that the variation associated with probe size is smaller than that observed across noise-strength ranges. Because this comparison covers only three small systems and the feature dimension grows with qubit count, it should be interpreted as a local sensitivity analysis rather than evidence of scalability.

\subsection{Limitations}

The current framework has several limitations. First, the main evaluation is restricted to simulated 3-qubit probe circuits. The additional 2- and 4-qubit comparison covers too narrow a range to establish scalability, and the growing observable and feature sets may increase measurement and training costs at larger sizes. Second, the pipeline performs discrete noise-model classification rather than continuous noise-parameter estimation. Third, all experiments are conducted in simulation and do not capture hardware-specific effects such as calibration drift, crosstalk, and correlated noise; the reported performance therefore does not establish generalization to physical devices. Finally, the current probe circuits may not explore a sufficiently diverse set of states to distinguish all ten noise models reliably.

\section{Conclusion and Future Work}\label{sec:conclusion}

This paper presented a shadow-based noise fingerprinting pipeline for classifying ten simulated quantum noise models using a fixed set of 3-qubit probe circuits. The three classifiers achieved broadly comparable performance, with the tree-based classifiers slightly ahead in the reported runs. Accuracy depended more strongly on the noise-strength range than on the nearby probe sizes examined. A limited comparison with analogous 2- and 4-qubit probes showed similar accuracy across these nearby sizes, but the range is too small to support a scalability claim. Future work will evaluate the framework on larger systems; extend it to continuous noise-parameter estimation; test it on physical hardware under finite-shot variation; and develop probe designs that better separate physically overlapping channels.

\bibliographystyle{IEEEtran}
\bibliography{refs}

\end{document}